# Understanding the Knowledge Sharing Behaviors of Library Professionals in South Asia

Abinash Deka
*Central University of Himachal Pradesh*, abinashdeka0@gmail.com

Subaveerapandiyan A
*Regional Institute of Education Mysore*, subaveerapandiyan@gmail.com

# Understanding the Knowledge Sharing Behaviors of Library Professionals in South Asia

**Abinash Deka**
Former Student
Master of Library and Information Science
Central University of Himachal Pradesh
Email: abinashdeka0@gmail.com

**Subaveerapandiyan A**
Professional Assistant (Library)
Regional Institute of Education Mysore,
India
Email: subaveerapandiyan@gmail.com

**Abstract**
   The present study aim is to know the information professionals/library professional's knowledge sharing behaviours and attitudes among the institutes. This study investigated six countries' library professionals: Bangladesh, Bhutan, India, Nepal, Pakistan, and Sri Lanka. The study discussed knowledge sharing behaviour, technological equipment used for knowledge management and disseminating the sources of knowledge; academic social networking sites used for sharing the information and knowledge as well as challenges in knowledge management faced by the librarians examined in detail. The implication of the study highlighted the various areas of knowledge management such as training, budget, lack of staff and reward.

1. **Introduction**
   In this 21$^{st}$ century, knowledge is an indispensable and has become a library that plays a vital role in knowledge and resource sharing. Knowledge sharing is one of the challenging processes for knowledge managers and knowledge center's between the user community. Knowledge sharing is not an effortless task because of various reasons behind resource sharing, such as legal issues, inadequate resource management, and distribution of knowledge resources. Resource managers have to follow and obey the author(s), publisher(s), policies, government guidelines, and all that. Knowledge sharing is essential in this present scenario; even a single library cannot provide various resources to the user demands, so consortium is more critical within the institutes. Without any expectation, the government has to support and promote the library consortia site. Knowledge is shareable with anyone in any place based on their need; anyone can provide knowledge if they are specialized in their field of work. The success of the library is sharing information and knowledge with others. Already experienced coding knowledge is called explicit knowledge; maybe it is any medium of multimedia format. Knowledge sharing disseminates the knowledge from one to another through multi-directional instead of uni-directional, group of people, or a specific community. Librarians have to know the expectations of their user's needs, interests, and specialized areas. Academic institutes and R&D institutes are creating more knowledge resources, and libraries are disseminating a wide range of knowledge resources.

## 2. Aim and Objectives of the study

- ❖ To identify the South Asian Librarian's familiarity of knowledge sharing
- ❖ To know the preparedness of knowledge sharing with others
- ❖ To examine information systems in facilitating the knowledge sharing
- ❖ To know the Knowledge sharing behavior among staff

## 3. Review of Literature

Parirokh & Farhad (2008) performed a study and found out that most librarians used formal and informal (face-to-face) communication to capture information sources. In addition, and simultaneously, some librarians communicate with other libraries as their information sources. Further, it also found the issues mentioned; perhaps it designed most current information technologies in libraries to perform specific functions rather than facilitate an organizational process. Therefore, the study suggested creating a knowledge management unit or officer who would enhance the knowledge-sharing activities. Appropriate ICT infrastructures are also highly recommended in academic libraries to facilitate specific knowledge management policies and improve the knowledge-sharing capabilities of librarians. In addition, they must provide various communication channels for librarians, enhancing both efficiency and effectiveness in communication and knowledge sharing activities.

Variant & Dyah (2013) found that knowledge sharing did not formally adopt in many libraries in Surabaya; only a few libraries have implemented it. The study discovered that the information communication infrastructure of libraries in Surabaya is fundamental, such as discussion rooms with computers and LCD projectors. Some libraries support knowledge sharing, but there is still a need for applications that promote collaboration virtually. The libraries lack the application of reward systems or incentives for staff who have been contributing to knowledge sharing; it triggers teams to reduce contribution and intention to knowledge sharing through forums. The libraries also lack knowledge reuse and open access maximally. The study suggested that the libraries need to be more severe in planning the knowledge-sharing strategy following the intended goal. The libraries should encourage the creation of knowledge and provide access to this knowledge for future use.

Awodoyin et al. (2016) carried out 12 selected academic libraries in Ogun State, Nigeria; they observed that most librarians, 82.9% preferred face-to-face interaction and mobile phones as knowledge-sharing channels. The study reveals that e-mail and newsletters are frequently used by 76.1% of librarians. The least knowledge-sharing channels used by the librarians were library blog, library portal 44.4%, and social media such as Facebook, Twitter, and Yahoo Messenger. The study also found significant problems that work against effective knowledge sharing utilizing librarians' lack of understanding on how to share knowledge 82.9% effectively, lack of social networking skills 69.2%, inability to use modern technology, and failure to appreciate the values of knowledge sharing 62.4%. Similarly, 65.8% of the respondents did not support that knowledge sharing depended on technology. The study and the findings suggested that the outcome of seminar and conference participation is not enough for librarians; instead, the library management should make a routine for open interaction between librarians within the library or outside the library to generate innovative ideas to help reshape the library.

Khan & Ali (2019) carried out a study on Indian academic library professionals perceived knowledge sharing as an exchange of one individual to another or group of individuals, such as documents, reports, manuals, meeting minutes or sharing their ideas, experiences, skills with the other staff. The study found that library professionals in Indian academic libraries have a positive attitude towards knowledge sharing. This study highlighted that in developing countries like India, libraries are still functioning through traditional methods, although few academic libraries have advanced ICT technologies. The significant barriers found in this study were lack of trust, personal animosity, technological support, nepotism, and cronyism at the workplace.

Kaffashan et al. (2020) found that various factors directly or indirectly influence librarian's knowledge sharing behavior. They are organizational climate, subjective norms, leadership empowerment, attitude, motivational drives, intention, and knowledge sharing, reducing or increasing their knowledge sharing. This study also encourages library managers to search for ways to improve current organizational conditions. Besides this, the study has also proposed a novel approach to enhance librarian's knowledge sharing behavior based on hypotheses of the theory of reasoned action (TRA), which analyses the direct and indirect effect of organizational factors on elements of the TRA.

Ahmed et al. (2020) have examined the six dimensions of knowledge sharing: innovation, collaboration, communication channels, trust, loyalty, and ethics. They found that organizational satisfaction is an important fact that enables communication and dedication among library professionals towards knowledge sharing. It reveals a connection between corporate culture and knowledge sharing factors, i.e., organizational satisfaction, support, and good leadership promote loyalty among library professionals. The study shows that employees prefer to work in an enjoyable environment and under reasonable supervision despite better opportunities in other libraries. They also willingly and openly express their expertise with their colleagues through presentation or groupware, or intranet. This study also suggests the library administrators be cautious towards the three factors of organizational culture: employees' satisfaction, good leadership, and organizational support. These factors play a significant role in growing the quality of services and improving library professionals' performance.

4. Method

This study applied a descriptive quantitative method to collect the data. The present study is based on primary data. For collecting the data mail IDs are collected from their official institute library websites. The structured questionnaires were prepared and distributed the questionnaires by email. The questionnaire comprises two parts—Part-I socio-demographic details of the respondents and Part-II Knowledge sharing behaviors of LIS professionals. The participants belonged to the Six South Asian Countries higher education institutes academic library professionals. The survey population comprised 175 respondents from Bangladesh, Bhutan, India, Nepal, Pakistan, Sri Lankan countries; after collecting the data, appropriate statistical analytical tools were used and analyzed. SPSS statistical tools were used and one-way ANOVA was performed to check any significant difference between the means of two or more groups ($p > 0.05$).

## 5. Limitations of the Study

The present study is conducted to identify and evaluate the level of knowledge sharing behavior of library professionals. The study will help understand the library professionals' knowledge sharing behavior and awareness and how it will affect the organization, librarian's community, and the surrounding population. The scope is limited to library professionals working in various university libraries in Bangladesh, Bhutan, India, Nepal, Pakistan, Sri Lankan countries excluding Afghanistan and Maldives because of time and other limitations. 175 library professionals from various academic libraries of the South-Asia region were involved in giving an idea about the overall knowledge sharing behavior of the academic librarians.

## 6. Data Analysis and Findings

**Table 1. Demographic Distribution of Respondents**

| Type | Division | Frequency | Percentage |
|---|---|---|---|
| Country | Bangladesh | 22 | 12.6 |
| | Bhutan | 7 | 4 |
| | India | 107 | 61.1 |
| | Nepal | 1 | 0.6 |
| | Pakistan | 27 | 15.4 |
| | Sri Lanka | 11 | 6.3 |
| Gender | Male | 117 | 66.9 |
| | Female | 58 | 33.1 |
| Age | 20-25 | 6 | 3.4 |
| | 26-30 | 20 | 11.4 |
| | 31-35 | 30 | 17.1 |
| | 36-45 | 67 | 38.3 |
| | 46-55 | 41 | 23.5 |
| | Above 55 | 11 | 6.3 |
| Qualifications | Certificate | 2 | 1.1 |
| | Diploma | 9 | 5.1 |
| | Bachelor's degree | 6 | 3.5 |
| | Masters' Degree | 87 | 49.7 |
| | M.Phil. | 14 | 8 |
| | PhD | 54 | 30.9 |
| | Not Qualified | 3 | 1.7 |
| **Designation** | Librarian | 66 | 37.7 |
|  | Deputy Librarian | 9 | 5.1 |
|  | Assistant Librarian | 49 | 28 |
|  | Library Assistant | 36 | 20.6 |
|  | Others | 15 | 8.6 |
| **Working Experience** | Less than 1 year | 3 | 1.7 |
|  | 2 to 5 | 29 | 16.5 |
|  | 6 to 10 | 26 | 14.9 |
|  | 11 to 15 | 40 | 22.9 |
|  | More than 15 | 77 | 44 |
| **Total** |  | 175 | 100 |

Results in the above table 1 show that, out of 175 respondents interviewed in South Asia, the majority are from India 61.1% of respondents, 66.9% of male respondents, 38.3% of respondents fall within the 36-45 age group. Designation wise more respondents are Librarians (37.7%). Educational qualifications wise, 49.7% of respondents have a master's degree as the highest education qualification, followed by those with a PhD 30.9%. Experience wise vast respondents were 44% with over 15 years of work experience in their respective fields.

**Table 2: Knowledge Sharing Awareness**

| Knowledge sharing awareness | Respondents | Percentage |
|---|---|---|
| Excellent | 66 | 37.7 |
| Good | 95 | 54.3 |
| Fair | 14 | 8 |
| Poor | 0 | 0 |
| Very poor | 0 | 0 |

**Fig 1: Knowledge Sharing Awareness of the respondents**

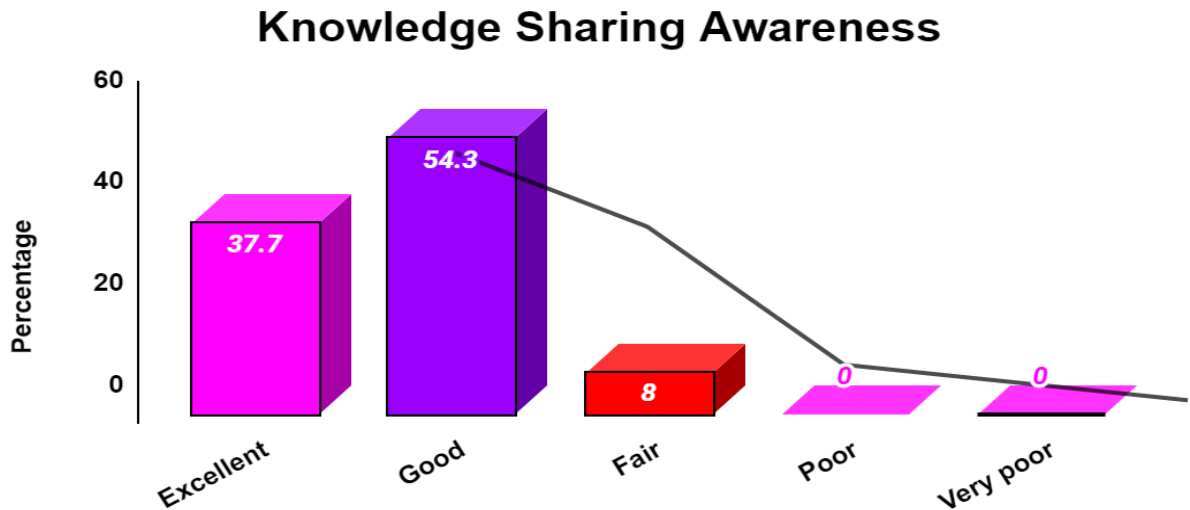

Table 2 and Figure 1 represents the knowledge sharing awareness of the respondents. The majority, 54.3% of respondents, have *'good'* knowledge sharing awareness, and 37.7% have *'excellent'* knowledge sharing awareness. Likewise, 8% of the respondents have *'fair'* knowledge sharing awareness.

**Fig 2: Willing to Share Knowledge of respondents**

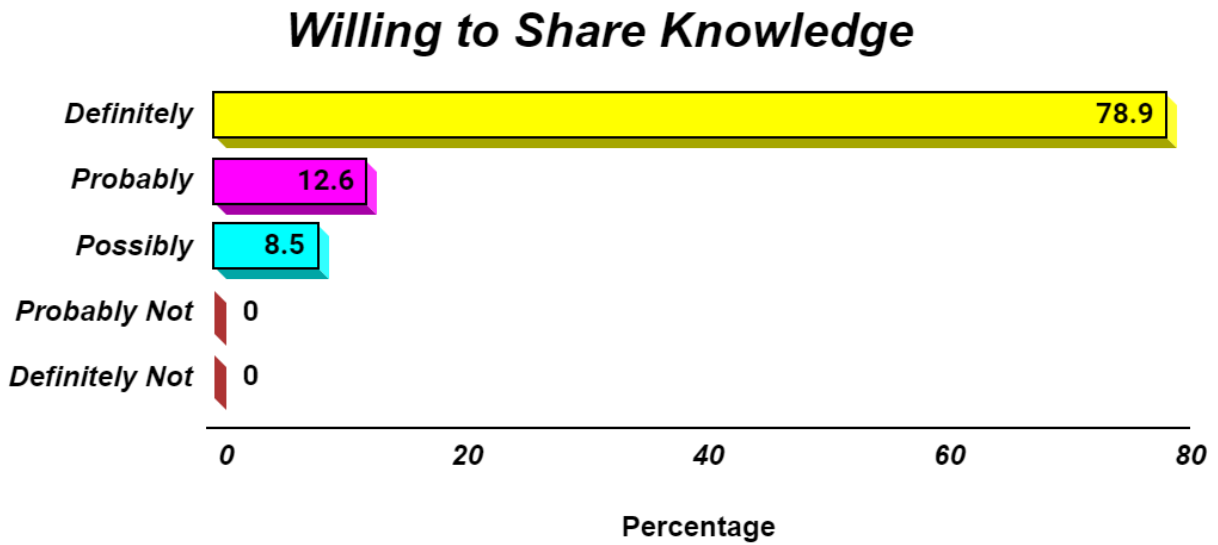

Figure 2 presents the respondent's readiness of one's own free will to share knowledge. It was found that most of the respondents, 78.9%, definitely share their knowledge, while 12.6% of the respondents may probably share knowledge. In contrast, 8.5% of the respondents possibly share knowledge willingly.

**Fig 3: Sources for acquiring knowledge**

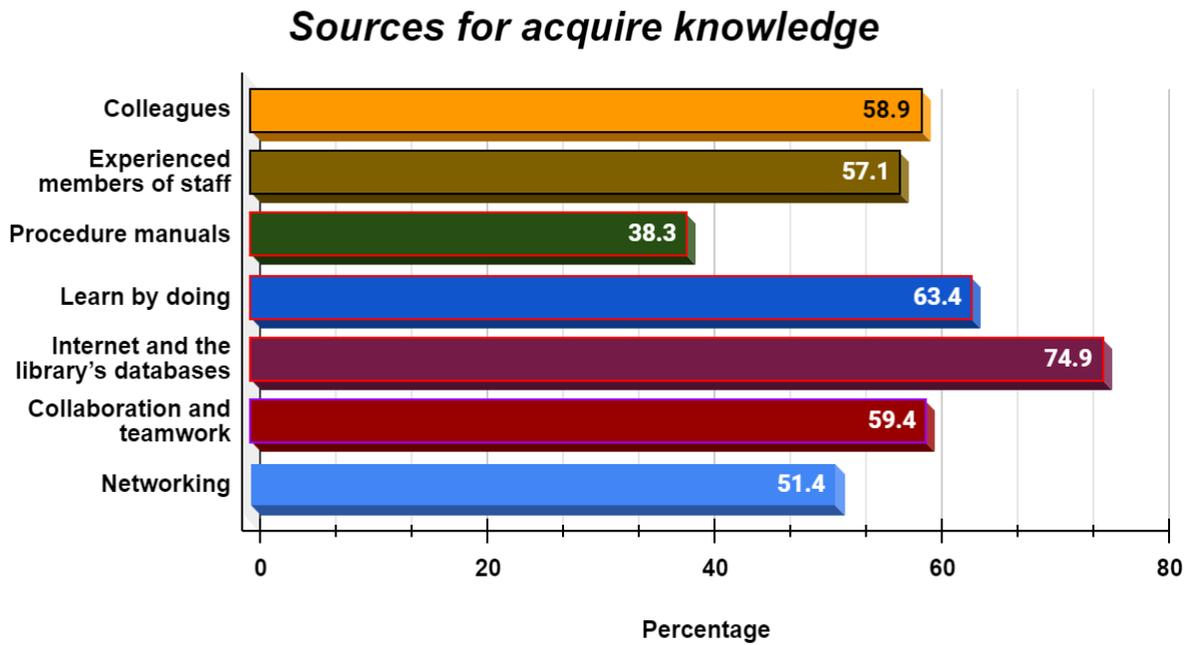

The above figure 3 shows the sources through which the respondents gain knowledge. About 74.9% of the respondents noted they gain knowledge from *'internet and other library databases'*, and 63.4% of respondents gain knowledge through *'learn by doing'* their respective work in the libraries. Also, the study findings show that 59.4% of the respondents get knowledge through *'collaboration and teamwork'*. Similarly, 58.9% of the respondents gain knowledge from *'colleagues'*, and 57.1% of the respondents gain or gain knowledge from *'experienced staff members'* in their respective organizations. Likewise, 51.4% of the respondents gain knowledge with *'networking'* activities, and 38.3% of the respondents receive knowledge from *'procedure manuals.*

**Table 3. Usage of technology in Knowledge management/Knowledge Sharing**

| Technology for Knowledge Sharing | Very Important | Important | Moderately Important | Slightly Important | Unimportant |
|---|---|---|---|---|---|
| Internet/ Intranet/Extranet | 139 (79.4%) | 30 (17.1%) | 5 (2.9%) | 1 (0.6%) | 0 |
| E-mail/Group mail | 113 (64.5%) | 53 (30.3%) | 5 (2.9%) | 4 (2.3%) | 0 |
| Video conferencing/ Teleconferencing/ Video sharing | 65 (37.1%) | 83 (47.5%) | 20 (11.4%) | 7 (4%) | 0 |
| Storytelling | 19 (10.9%) | 66 (37.7%) | 54 (30.9%) | 28 (16%) | 8 (4.5%) |
| Data management system | 95 (54.3%) | 65 (37.1%) | 13 (7.5%) | 2 (1.1%) | 0 |
| Data support system | 83 (47.4%) | 71 (40.6%) | 17 (9.7%) | 4 (2.3%) | 0 |
| Content Management | 90 (51.4%) | 69 (39.5%) | 13 (7.4%) | 2 (1.1%) | 1 (0.6%) |
| Knowledge Portals | 86 (49.1%) | 74 (42.3%) | 13 (7.4%) | 1 (0.6%) | 1 (0.6%) |
| Instant messaging/Online chatting | 59 (33.7%) | 76 (43.5%) | 21 (12%) | 13 (7.4%) | 6 (3.4%) |
| Wikis/ Groupware/Online discussion forums | 52 (29.7%) | 69 (39.4%) | 47 (26.9%) | 7 (4%) | 0 |
| Blogs/ YouTube/Facebook/ Twitter | 56 (32%) | 70 (40%) | 38 (21.7%) | 7 (4%) | 4 (2.3%) |

Above table 3 shows the usage of technology in knowledge management and knowledge sharing. It was found that 79.4%, most respondents, considered the usage of *'Internet, Intranet and extranet'* as very important for knowledge management and knowledge sharing. In contrast, 64.5% of respondents considered the usage of *'e-mail/group mail'* as very important. Also, 47.5% of respondents indicated the usage of *'video conferencing and teleconferencing'* as necessary. Furthermore, 37.7% of respondents considered *'storytelling'* as important in knowledge management and knowledge sharing. Further, 54.3% considered the *'data management system'* very important, and 47.4% considered the *'database support system'* as very important for knowledge management and knowledge sharing. Likewise, 51.4% of respondents considered *'content management'* very important, and 49.1% considered *'knowledge portals'* vital. Likewise, 43.5% considered *'instant messaging and online chatting'* as necessary. Further, 39.4% consider using *'wikis, groupware and online discussion forums'* as necessary, and 40% of respondents considered *'blogs, YouTube, Facebook and Twitter'* as important ways for knowledge management and knowledge sharing.

**Fig 4: Academic Social Networking Sites for Knowledge Sharing**

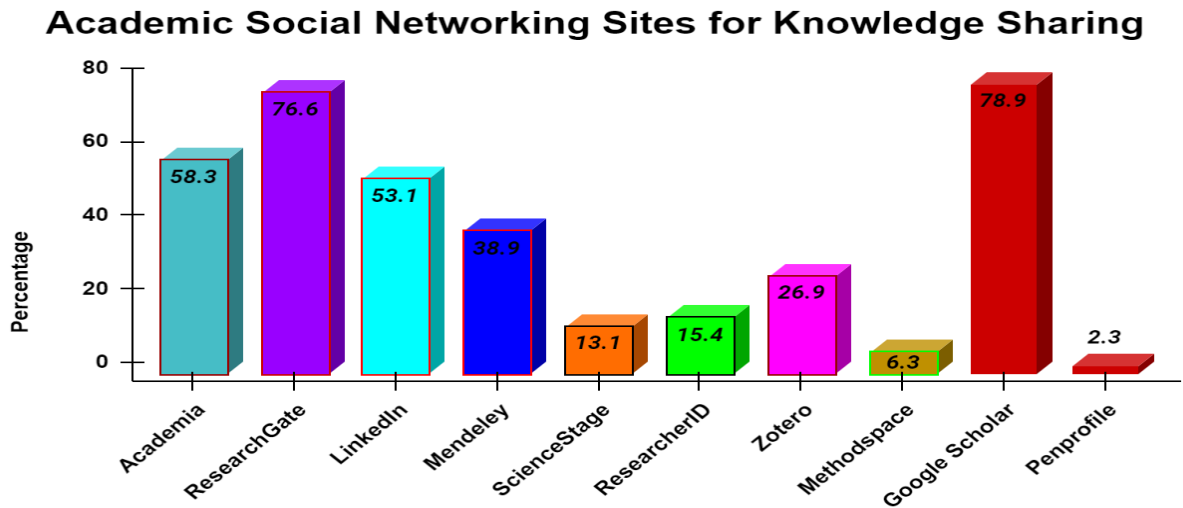

The above figure 4 shows the usage of various academic and social networking sites for knowledge sharing. It was found that the majority 78.9% of the respondents used *'Google Scholar'*, followed by 76.6%, preferred *'ResearchGate'*. 58.3% considered *'Academia'* for knowledge sharing, while 53.1% considered usage of *'LinkedIn'* for knowledge sharing. Similarly, 38.9% of the respondents considered *'Mendeley'* and 26.9% of respondents believed *'Zotero'* for 'knowledge sharing'. Also, 15.4% considered *'ResearcherID'* for knowledge sharing and 13.1% considered *'ScienceStage'* for knowledge sharing. Only 2.3% of respondents considered *'Penprofile'* for 'knowledge sharing'.

**Table 4. Way to encouraging staff members to share their knowledge**

| Way to encouraging | SA | A | N | D | SD | Average |
|---|---|---|---|---|---|---|
| Encouraged with incentives | 57 (32.6%) | 66 (37.7%) | 28 (16%) | 22 (12.6%) | 2 (1.1%) | 3.88 ± 1.04 [a] |
| Encouraged to publishing scholarly articles | 72 (41.1%) | 84 (48%) | 18 (10.3%) | 1 (0.6%) | 0 | 4.29 ± 0.67 [a,b] |
| Encouraged to become members of professional bodies | 64 (36.6%) | 102 (58.3%) | 9 (5.1%) | 0 | 0 | 4.31 ± 0.56 [a,b] |
| Encouraged to attend/ give guest lectures | 62 (35.4%) | 99 (56.6%) | 12 (6.9%) | 2 (1.1%) | 0 | 4.26 ± 0.63 [a,b] |
| Encouraged to conduct conferences, seminars, webinars & workshops | 74 (42.3%) | 92 (52.6%) | 9 (5.1%) | 0 | 0 | 4.37 ± 0.58 [a,b] |
| Encouraged to take part in conferences, seminars, webinars & workshops | 93 (53.1%) | 79 (45.2%) | 3 (1.7%) | 0 | 0 | 4.51 ± 0.53 [b] |
| Institutions should have a policy of encouraging the innovative initiatives of their employees | 79 (45.2%) | 91 (52%) | 5 (2.9%) | 0 | 0 | 4.42 ± 0.55 [a,b] |
| Regular email shots and weekly newsletters | 51 (29.1%) | 102 (58.3%) | 18 (10.3%) | 3 (1.7%) | 1 (0.6%) | 4.13 ± 0.7 [a,b] |
| KM is building a culture of knowledge learning, sharing, & development | 72 (41.1%) | 89 (50.9%) | 12 (6.9%) | 2 (1.1%) | 0 | 4.32 ± 0.65 [a,b] |
| Making information available at all levels | 70 (40%) | 93 (53.1%) | 11 (6.3%) | 1 (0.6%) | 0 | 4.32 ± 0.61 [a,b] |
| Need to conduct effective education & training to develop a knowledge-sharing | 81 (46.3%) | 87 (49.7%) | 4 (2.3%) | 3 (1.7%) | 0 | 4.40 ± 0.62 [a,b] |

Scale Used: 1 Strongly disagree; 2 Disagree; 3 Neutral; 4 Agree; 5 Strongly agree; different letter suffices denote significant ($p<0.05$) variations in 'average'.

Table 4 shows the various ways to encourage staff members to share their knowledge. The majority, 37.7% of the respondents, agreed that it is 'encouraged with incentives'. Likewise, 48% agreed that it is 'encouraged to publish scholarly articles'. Similarly, 58.3% of respondents agreed that 'encouraged to become members of professional bodies' and 'regular email shots and weekly newsletters'. Additionally, 56.6% agreed that 'encouraged to attend/give guest lectures' will encourage staff knowledge sharing behavior. Also, 52.6% agree with 'encouraged to conduct conference, seminars, webinars and workshops', followed by 53.1% who strongly agree that 'encouraged to participate in conference, seminars, webinars and workshops. 52% agreed that 'institutions should have a policy of encouraging the innovative initiatives of their employees', and 41.1% strongly agreed that 'knowledge management is building a culture of knowledge learning, sharing and development'. In addition, 53.1% agreed that 'making information available at all levels' may encourage knowledge sharing behavior. Also, 49.7% agreed that 'need to conduct effective education and training to develop a knowledge-sharing culture in the organization'.

**Table 5. Challenges of Knowledge Management (KM)**

| Challenges of KM | SA | A | N | D | SD | Average |
|---|---|---|---|---|---|---|
| The unfavorable organizational culture that impedes knowledge sharing behavior | 34 (19.4%) | 85 (48.6%) | 31 (17.7%) | 23 (13.2%) | 2 (1.1%) | $3.72 \pm 0.96^a$ |
| Lack of relevant training | 55 (31.4%) | 86 (49.1%) | 25 (14.3%) | 8 (4.6%) | 1 (0.6%) | $4.06 \pm 0.83^b$ |
| Lack of clearly defined guidelines on knowledge management implementation | 49 (28%) | 93 (53.1%) | 26 (14.9%) | 7 (4%) | 0 | $4.05 \pm 0.76^b$ |
| Insufficient and inappropriate technological systems | 53 (30.3%) | 77 (44%) | 22 (12.6%) | 20 (11.4%) | 3 (1.7%) | $3.89 \pm 1.01^{a,b}$ |
| Librarians lack expertise in knowledge management | 30 (17.1%) | 82 (46.9%) | 38 (21.7%) | 18 (10.3%) | 7 (4%) | $3.62 \pm 1.01^a$ |
| Lack of organizational leadership commitment | 37 (21.1%) | 82 (46.9%) | 36 (20.6%) | 20 (11.4%) | 0 | $3.77 \pm 0.91^a$ |
| Lack of reward system and incentives | 41 (23.4%) | 80 (45.7%) | 35 (20%) | 15 (8.6%) | 4 (2.3%) | $3.79 \pm 0.97^a$ |
| Lack of awareness of knowledge management concepts | 46 (26.3%) | 87 (49.7%) | 25 (14.2%) | 12 (6.9%) | 5 (2.9%) | $3.89 \pm 0.96^{a,b}$ |
| Limited budgets | 60 (34.3%) | 70 (40%) | 28 (16%) | 9 (5.1%) | 8 (4.6%) | $3.94 \pm 1.05^b$ |

Scale Used: 1 Strongly disagree; 2 Disagree; 3 Neutral; 4 Agree; 5 Strongly agree; different letter suffices denote significant (p<0.05) variations in 'average'.

The above-given Table 5 shows the detailed analysis of various challenges of Knowledge Management (KM). The topmost among the challenges found was *'lack of clearly defined guidelines on knowledge management implementation'* 53.1% majority of respondents agreeing with it. Others in ascending order are, *'Lack of awareness of knowledge management concepts'*, 49.7% of respondents agreed. *Furthermore, 'lack of relevant training'* which 49.1% of respondents agreed, followed by *'the unfavorable organizational culture that impedes knowledge sharing behavior'* which 48.6% of respondents agreed, equally 46.9% of respondents agreed that *'Librarians lack of expertise in knowledge management'* and *'lack of organizational leadership commitment'* as the challenges involved during knowledge management (KM). Additionally, 44% of the respondents agreed that *'Insufficient and inappropriate technological systems'* are challenging in knowledge management. Finally, 40% of the respondents agreed that *'Limited budgets'* can be a challenge in knowledge management (KM).

**Table 6: Barriers in knowledge sharing**

| Barriers in Knowledge sharing | SA | A | N | D | SD | Average |
|---|---|---|---|---|---|---|
| Lack of documents | 18 (10.3%) | 56 (32%) | 33 (18.9%) | 59 (33.7%) | 9 (5.1%) | $2.91 \pm 1.12^a$ |
| Lack of time | 20 (11.4%) | 53 (30.3%) | 34 (19.4%) | 57 (32.6%) | 11 (6.3%) | $2.92 \pm 1.15^a$ |
| Ethical Issues | 11 (6.3%) | 47 (26.9%) | 37 (21.1%) | 65 (37.1%) | 15 (8.6%) | $3.14 \pm 1.10^b$ |
| Legal Issues | 17 (9.7%) | 27 (15.4%) | 46 (26.3%) | 68 (38.9%) | 17 (9.7%) | $3.18 \pm 1.13^b$ |
| Lack of staff commitment | 17 (9.7%) | 29 (16.6%) | 42 (24%) | 64 (36.6%) | 23 (13.1%) | $3.26 \pm 1.17^b$ |
| Lack of staff | 15 (8.6%) | 32 (18.3%) | 25 (14.2%) | 61 (34.8%) | 42 (24%) | $3.47 \pm 1.27^b$ |
| Lack of knowledge | 14 (8%) | 43 (24.6%) | 42 (24%) | 56 (32%) | 20 (11.4%) | $3.14 \pm 1.15^b$ |

Scale Used: 1 Strongly disagree; 2 Disagree; 3 Neutral; 4 Agree; 5 Strongly agree; different letter suffices denote significant (p<0.05) variations in 'average'.

Table 6 shows a detailed analysis of various barriers to knowledge sharing. It shows that the majority 38.9% of respondents *'Agreed'* with *'with legal Issues'* as the barrier in knowledge sharing, followed by 37% of respondents, *'Agreed'* with *'Ethical issues'*. Almost equally, 36.6% of respondents *'Agreed'* with *'Lack of staff commitment'*. Further, 34.8% of respondents *'Agreed'* with *'Lack of staff'*. While 33.7% of respondents *'Agreed'* with *'Lack of documents'*, 32.6% of respondents *'Agreed'* with *'Lack of time'* as the barriers in knowledge sharing. Moreover, 32% of respondents *'Agreed'* with *'Lack of knowledge'* as the barrier in knowledge sharing.

## 7. Findings and Conclusion

Knowledge sharing awareness and knowledge sharing behavior play an essential role in creating new knowledge in every growing organization. Knowledge sharing is vital since it facilitates decision-making capabilities within organizations. Knowledge sharing also improves performance at work, effectiveness at work, and skills. This study has revealed that most of the library professionals of the various academic institutions of the South-Asia region have a good level of knowledge sharing awareness and take part in knowledge sharing activities through various mediums such as library databases and various other academic networking sites such as knowledge portals, conferences, webinars. Email, group mail, internet and social networking sites such as blogs, YouTube, Facebook, Twitter. However, a few library professionals disagree entirely with instant messaging and online chatting tools and social networking sites in knowledge management and knowledge sharing. Further, this study also revealed various ways to encourage library professionals to participate in knowledge sharing activities, such as encouragement with incentives, encourage professionals to participate in scholarly communication, conduct seminars and webinars frequently, and encourage the innovative initiatives of the employees. The study also revealed various challenges involved in knowledge sharing such as lack of proper knowledge management, lack of proper training of the staff, unfavorable organizational culture, insufficient ICT infrastructures, lack of reward system and incentives, lack of organizational leadership commitment and limited budgets.

The study revealed that most of the library professionals in the South-Asia region engaged with all forms of knowledge sharing activities in all ways such as social media, academic networking sites, library databases, knowledge portals, conferences, online chat, Email, Internet, Intranet and group mail, Experienced members of staff and collaboration and teamwork. However, they faced limitations such as lack of documents, time, ethical and legal issues, lack of staff and commitment, and lack of knowledge.

This study showed the various ways to motivate library professionals to participate in knowledge sharing, such as encouraging incentives, publication of research articles in reputed journals, and attending guest lectures/special lectures and audio-visual presentations. However, this study revealed the various challenges of knowledge management such as lack of clearly defined guidelines on knowledge management and implementation, lack of organizational leadership commitment, lack of expertise in knowledge management, lack of relevant training, lack of reward system and incentives, lack of awareness in knowledge management concepts, insufficient and inappropriate technological systems, limited budgets and unfavorable organizational culture.